\documentclass[journal,draftcls,onecolumn,10pt,twoside]{IEEEtranTMBMC}

\usepackage{amsmath}
\usepackage{color}
\usepackage{array}
\usepackage{stfloats}
\usepackage{amsthm} 
\usepackage{amssymb}
\usepackage{float}
\usepackage{nccmath}
\usepackage{graphicx}
\usepackage[linesnumbered,ruled]{algorithm2e}
\usepackage{setspace}
\usepackage{booktabs}
\usepackage{subcaption}
\usepackage{mwe}
\usepackage{cite}
\usepackage{amsmath,amssymb,amsfonts}
\usepackage{graphicx}
\usepackage{textcomp}
\usepackage{xcolor}
\usepackage{tikz}

\usepackage{listings}
\usepackage{xcolor}
\usepackage[utf8]{inputenc}

\usepackage{algorithmicx}
\usepackage{xspace}
\usepackage{physics}

\definecolor{codegreen}{rgb}{0,0.6,0}
\definecolor{codegray}{rgb}{0.5,0.5,0.5}
\definecolor{codepurple}{rgb}{0.58,0,0.82}
\definecolor{backcolour}{rgb}{0.95,0.95,0.92}

\lstdefinestyle{mystyle}{
    language=Python,
    backgroundcolor=\color{backcolour},   
    commentstyle=\color{codepurple}\bfseries,
    keywordstyle=\color{codegreen}\bfseries,
    numberstyle=\tiny\color{codegray},
    stringstyle=\color{red},
    basicstyle=\ttfamily\footnotesize, 
    breakatwhitespace=false,         
    breaklines=true,                 
    captionpos=b,                    
    keepspaces=true,                 
    numbers=right,                    
    numbersep=1pt,                  
    showspaces=false,                
    showstringspaces=false,
    showtabs=false,                  
    tabsize=2,
    morekeywords={yield},
    emph ={problem_op, mixing_op},
    emphstyle=\color{blue},
    deletekeywords=[2]{compile},
}

\renewcommand\baselinestretch{1.00}

\begin{document}
\title{Quantum Approximation for Wireless Scheduling}

\author{Jaeho Choi, Seunghyeok Oh, and Joongheon Kim
    \thanks{This research was supported by National Research Foundation of Korea (2019M3E4A1080391, Development of Quantum Deep Reinforcement Learning Algorithm using QAOA).}
	\thanks{J. Choi is with the School of Computer Science and Engineering, Chung-Ang University, Seoul, Korea e-mail: jaehochoi2019@gmail.com.}
	\thanks{S. Oh is with the Department of Physics, Chung-Ang University, Seoul, Korea e-mail: seunghyeokoh2019@gmail.com.}
	\thanks{J. Kim is with the School of Electrical Engineering, Korea University, Seoul, Korea e-mail: joongheon@korea.ac.kr.}
	\thanks{J. Kim is a corresponding author of this paper.}
}

\maketitle

\begin{abstract}
This paper proposes a quantum approximate optimization algorithm (QAOA) method for wireless scheduling problems. 
The QAOA is one of the promising hybrid quantum-classical algorithms for many applications and it provides highly accurate optimization solutions in NP-hard problems.
QAOA maps the given problems into Hilbert spaces, and then it generates Hamiltonian for the given objectives and constraints. 
Then, QAOA finds proper parameters from classical optimization approaches in order to optimize the expectation value of generated Hamiltonian.
Based on the parameters, the optimal solution to the given problem can be obtained from the optimum of the expectation value of Hamiltonian. 
Inspired by QAOA, a quantum approximate optimization for scheduling (QAOS) algorithm is proposed. 
First of all, this paper formulates a wireless scheduling problem using maximum weight independent set (MWIS).
Then, for the given MWIS, the proposed QAOS designs the Hamiltonian of the problem. 
After that, the iterative QAOS sequence solves the wireless scheduling problem. 
This paper verifies the novelty of the proposed QAOS via simulations implemented by Cirq and TensorFlow-Quantum.
\end{abstract}
\begin{keywords}
Quantum Approximate Optimization Algorithm (QAOA), Maximum Weight Independent Set (MWIS), NP-Hard, Wireless Scheduling, Quantum Application
\end{keywords}
\IEEEpeerreviewmaketitle

\section{Introduction}
Nowadays, quantum computing and communications have received a lot of attention from academia and industry research communities. 
In particular, quantum computing based NP-hard problem solving is of great interest~\cite{arute, 01farhi, kandala, troyer}.
Among them, quantum approximate optimization algorithm (QAOA) is one of the well-known quantum computing based optimization solvers, and it has been verified that the QAOA outperforms the others in many combinatorial problems which is closed related to wireless scheduling problems~\cite{farhi, preskill, 20choi, zhou, choi}. 
Based on this nature, it is obvious that quantum computing can be used for various communications applications~\cite{nawaz, tariq, viswanathan, tang}.

In this paper, a wireless scheduling problem is formulated with maximum weight independent set (MWIS) formulation where the weight is defined as the queue-backlog to be transmitted over wireless channels~\cite{16kim, basagni, paschalidis}.
According to the fact that the MWIS problem is an NP-hard, heuristic algorithms are desired, thus a QAOA application algorithm, quantum approximate optimization for scheduling (QAOS), is designed to solve MWIS-based wireless scheduling problems.

The proposed QAOS works as follows. 
First of all, the objective function and constraint functions are formulated for MWIS.
Next, the corresponding objective Hamiltonian and constraint Hamiltonian are designed which map the objective function and the constraint function, respectively; and then, the problem Hamiltonian which should be optimized is formulated as the form of linear combinations of the objective Hamiltonian and constraint Hamiltonian.
In addition, the mixing Hamiltonian is formulated using a Pauli-$X$ operator.
Based on the definitions of the problem Hamiltonian and the mixing Hamiltonian, two corresponding unitary operators, i.e., problem operator and mixing operator, can be defined, respectively; and then parameterized state can be generated by alternately applying the two unitary operators.
The sample solutions can be obtained by the measurement of the expectation value of problem Hamiltonian on the parameterized state, and the parameters can be optimized in a classical optimization loop.
Finally, the optimal solution of the MWIS problem can be obtained by the measurement of the expectation value of problem Hamiltonian on the state generated by optimal parameters.
As verified in performance evaluation, the QAOS outperforms the other algorithms, e.g., random search and greedy search. 

The rest of this paper is organized as follows.
Section~\ref{sec:2} presents preliminary knowledge.
Section~\ref{sec:3} introduces MWIS-based wireless scheduling modeling.
Section~\ref{sec:4} presents the details of proposed QAOS algorithm, and the performance is evaluated in Section~\ref{sec:5}. 
Finally, Section~\ref{sec:6} concludes.

\section{Preliminaries}\label{sec:2}
Prior to problem-modeling, this section briefly explains bra-ket notation and Quantum Approximate Optimization Algorithm (QAOA)~\cite{farhi}.

\subsection{Bra-ket Notation}
In quantum computing, the bra-ket notation is generally used to represent qubit states (or quantum states).
It is also called Dirac notation as well as the notation for observable vectors in Hilbert spaces.
A ket and a bra can represent the column and row vectors, respectively.
Thus, single qubit states, i.e., $\ket{0}$ and $\ket{1}$, are represented as follows:
%
\begin{equation}
    \ket{0} =    
    \begin{bmatrix}
    1\\
    0\\
    \end{bmatrix},
    \text{ and }
    \ket{1} =    
    \begin{bmatrix}
    0\\
    1\\
    \end{bmatrix}, 
\end{equation}
\begin{eqnarray}
\text{and also } \ket{0}={\bra{0}}^{\dagger}&=&
    \begin{bmatrix}
    1&0
    \end{bmatrix}^{\dagger}, \\
\ket{1}={\bra{1}}^{\dagger}&=&
    \begin{bmatrix}
    0&1
    \end{bmatrix}^{\dagger}. 
\end{eqnarray}    
Note that, $\dagger$ means Hermitian transpose.
Accordingly, the superposition state of a single qubit can be represented as follows:
\begin{equation}
    c_1\ket{0}+c_2\ket{1}=
    \begin{bmatrix}
    c_1\\
    c_2\\
    \end{bmatrix},
\end{equation}
where $c_1$ and $c_2$ are probability amplitudes that are complex numbers.

\subsection{Quantum Approximate Optimization Algorithm (QAOA)}
QAOA is one of the well-known noisy intermediate-scale quantum (NISQ) optimization algorithms to combat combinatorial problems~\cite{farhi, preskill, 20choi, zhou}.
QAOA formulates $H_P$ (i.e., problem Hamiltonian) and $H_M$ (i.e., mixing Hamiltonian) from the objective function $f(y)$; and then generates the parameterized states $\ket{\gamma, \beta}$ by alternately applying the $H_P$ and $H_M$ on initial state $\ket{s}$.
Here, $f(y)$, $H_P\ket{y}$, $H_M$, and $\ket{\gamma, \beta}$ are defined as follows:
%
\begin{eqnarray}
    f(y) &\triangleq&f(y_1,y_2,...,y_n), \\
    H_P\ket{y} &\triangleq&f(y)\ket{y}, \\
    H_M &\triangleq&\sum_{k=1}^{n} X_k, \label{eq:org_HM}\\
    \ket{\gamma, \beta} &\triangleq& e^{-i\beta_p H_M} e^{-i\gamma_p H_P} \cdots \nonumber \\
    & & e^{-i\beta_2 H_M} e^{-i\gamma_2 H_P}e^{-i\beta_1 H_M} e^{-i\gamma_1 H_P} \ket{s}, \label{eq:state}
\end{eqnarray}
where $n\in\mathbb{Z}^{+}$, $p\in\mathbb{Z}^{+}$, and $X_{k}$ is the Pauli-$X$ operator applying on the $k^{th}$ qubit.
 
In QAOA, through iterative measurement on $\ket{\gamma, \beta}$, the expectation value of $H_P$ should be taken, and then eventually, the samples of $f(y)$ should be computed as follows: 
%
\begin{equation}\label{exp_f}
    \expval{f(y)}_{\gamma, \beta} = \expval{H_P}{\gamma, \beta}.
\end{equation}
The optimal values of the parameters $\gamma$ and $\beta$ can be obtained by classical optimization methods, e.g., gradient descent~\cite{zinkevich}.
Therefore, the solution can be computed from~\eqref{exp_f} via the parameters obtained.
Eventually, QAOA is a hybrid quantum-classical optimization algorithm in which proper Hamiltonian design and discovery of good parameters in a classical optimization loop are key~\cite{hadfield, streif, wang}.

\section{Wireless Scheduling Modeling using Maximum Weight Independent Set (MWIS)}\label{sec:3}
Suppose a wireless network consists of the set of one-hop links~\cite{16kim}. 
For the scheduling, a conflict graph is organized where the set of vertices is (the links) and two vertices are connected by an edge if the corresponding links suffer from interference. 
The conflict graph can be formulated by its adjacency matrix, whose $\mathcal{E}_{(i,j)}$ are defined as follows:
\begin{equation}
    \mathcal{E}_{(i,j)} = \left\{
        \begin{array}{ll}
             1, & \text{if $l_{i}$ interferes with $l_{j}$ where }\\
              & \text{$l_{i}\in\mathcal{L}$, $l_{j}\in\mathcal{L}$, and $i\neq j$},\\
             0, & \text{otherwise}.
        \end{array}
    \right.
    \label{eq:adj-e}
\end{equation}

For wireless network scheduling, the objective is for finding the set of links (i.e., nodes of the conflict graph) where adjacent two connected links via edges cannot be simultaneously selected because the adjacent two connected links are interfering to each other.
This is equivalent to the case which maximizes the summation of weights of all possible independent sets in a given conflict graph. 
Thus, it is obvious that wireless network scheduling can be formulated with MWIS as follows:
\begin{eqnarray}
    \max: & & \sum_{\forall l_{k}\in\mathcal{L}} w_{k}\mathcal{I}_{k}, \label{obj_f} \\
    \text{s.t.} & & \mathcal{I}_{i}+\mathcal{I}_{j}+\mathcal{E}_{(i,j)}\leq 2, \forall l_{i}\in\mathcal{L}, \forall l_{j}\in\mathcal{L}, \label{constraint_begin} \\
    & & \mathcal{I}_{i}\in\{0,1\}, \forall l_{i}\in\mathcal{L}, \\
    \text{where} & & 
        \mathcal{I}_{i} = \left\{
        \begin{array}{ll}
             1, & \text{if $l_{i}$ is scheduled where $l_{i}\in\mathcal{L}$}, \\
             0, & \text{otherwise}.
        \end{array}
    \right. \label{constraint_end}
\end{eqnarray}
Here, $w_{k}$ is a positive integer weight at $\forall l_{k}\in\mathcal{L}$. 
The above formulation ensures that conflicting links are not scheduled simultaneously: If $\mathcal{E}_{(i,j)}=0$ (no edge between $l_{i}$ and $l_{j}$), then $\mathcal{I}_{i}+\mathcal{I}_{j}\leq 2$, i.e., both indicator functions can be $1$. 
In contrast, if $\mathcal{E}_{(i,j)}=1$, $\mathcal{I}_{i}+\mathcal{I}_{j}\leq 1$, i.e., at most one of the two indicators can be $1$.
In wireless communication research, the $w_{k}$ where $\forall l_{k}\in\mathcal{L}$ is usually considered as transmission queue-backlog at which should be processed when the link is scheduled.
More details are in~\cite{16kim}.

\section{Quantum Approximate Optimization for Scheduling (QAOS)}\label{sec:4}
In this section, Hamiltonians of QAOA are designed based on the scheduling model in Section~\ref{sec:3}; and then Quantum Approximate Optimization for Scheduling (QAOS) algorithm is proposed by applying the designed Hamiltonian to QAOA.

\subsection{Design the problem Hamiltonian}
The problem Hamiltonian $H_P$ is designed by a linear combination of the objective Hamiltonian $H_O$ and the constraint Hamiltonian $H_C$.
The objectives and constraints of the problem are contained by $H_O$ and $H_C$, respectively.

\subsubsection{Design the objective Hamiltonian} 
Suppose that a basic Boolean function $B_1(x)$ exists as follows:
%
\begin{equation}\label{eq:boolean_f}
    B_1(x)=x \text{ where } x\in \{0,1\}.
\end{equation}
Due to quantum Fourier expansion,~\eqref{eq:boolean_f} can be mapped to Boolean Hamiltonian $H_{B_1}$ where $I$ and $Z$ are an Identity operator and the Pauli-$Z$ operator, respectively~\cite{18hadfield}:
%
\begin{equation}\label{eq:boolean_H}
    H_{B_1}=\frac{1}{2}(I-Z).
\end{equation}

According to~\eqref{eq:boolean_f}--\eqref{eq:boolean_H}, the objective function~\eqref{obj_f} can be mapped to the following Hamiltonian:
%
\begin{equation}\label{H_O'}
    H_{O'}=\sum_{\forall l_{k}\in\mathcal{L}}  \frac{1}{2} w_k (I- Z_k),
\end{equation}
where $Z_{k}$ is the Pauli-$Z$ operator applying on $\mathcal{I}_{k}$.
The objective of the model is to maximize $H_{O'}$, thus the objective Hamiltonian $H_O$ which should be minimized is as follows:
%
\begin{equation}\label{eq:H_O}
    H_{O}=\sum_{\forall l_{k}\in\mathcal{L}}  \frac{1}{2}w_k Z_k.
\end{equation}

\subsubsection{Design the constraint Hamiltonian}
\begin{figure}[t]
\footnotesize
\centering
\begin{tikzpicture}
    [
    blacknode/.style={circle,draw=black!100,fill=black!100,thick,inner sep=5},
    whitenode/.style={circle,draw=black!100,fill=white!100,thick,inner sep=5}
    ]
    \node[blacknode] (n00) at (0,0) 
    {\textbf{\textcolor{white}{1}}};
    \node[blacknode] (n10) at (4,0) [label=right: \text{Case C: Both Scheduled}] 
    {\textbf{\textcolor{white}{1}}};
    \node[whitenode] (n01) at (0,1) 
    {\textbf{0}};
    \node[blacknode] (n11) at (4,1) [label=right: \text{Case B: 1 Node Scheduled}] 
    {\textbf{\textcolor{white}{1}}};
    \node[whitenode] (n02) at (0,2) 
    {\textbf{0}};
    \node[whitenode] (n12) at (4,2) [label=right: \text{Case A: Both Unscheduled}] 
    {\textbf{0}};
    \draw[-] (0.4,0) -- (3.6,0)
    node [below,align=center,midway]
    {
    \textbf{$E_C(N_i, N_j)$}
    };
    \draw[-] (0.4,1) -- (3.6,1)
    node [below,align=center,midway]
    {
    \textbf{$E_B(N_i, N_j)$}
    };
    \draw[-] (0.4,2) -- (3.6,2)
    node [below,align=center,midway]
    {
    \textbf{$E_A(N_i, N_j)$}
    };
\end{tikzpicture}
    \caption{The number of possible cases when a single edge exists between two nodes in the conflict graph. The scheduled and unscheduled nodes have states $\ket{1}$ and $\ket{0}$, respectively. $N_i$ and $N_j$ represent arbitrary nodes, and $E_A(N_i, N_j)$, $E_B(N_i, N_j)$, and $E_C(N_i, N_j)$ represent edges in each case.}
    \label{fig:edge-node}
\end{figure}
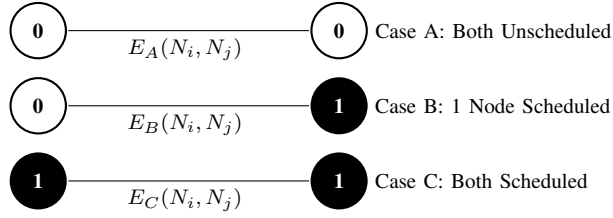
In the MWIS-based wireless scheduling problem, the banned condition is a case where both nodes of the conflict graph directly connected are scheduled, as shown in~\textit{Case C} of Fig.~\ref{fig:edge-node}.
If the weights of the $N_i$ and $N_j$ in \textit{Case C} are defined as $W_{N_i}$ and $W_{N_j}$ respectively; then the constraint function $C'(i,j)$, which computes banned conditions can be represented as follows:
%
\begin{equation}
    C'(i,j)=\sum_{i=1}^{n} \sum_{j=1}^{n}  (W_{N_i}+W_{N_j})\abs{E_C(N_i,N_j)} \text{ where } i>j.
\end{equation}
Here, $n$ is the number of nodes and $\abs{E_C(N_i,N_j)}$ is the number of $E_C(N_i,N_j)$; and $i>j$ is a condition to avoid duplication of the same edge.

According to~\eqref{eq:adj-e}--\eqref{constraint_end}, $C'(i,j)$ can be redefined to $C(i,j)$ with symbols in Section~\ref{sec:3} as follows:
%
\begin{eqnarray}\label{eq:C}
    C(i,j)&=&\sum_{{\forall l_{i}\in\mathcal{L}}} \sum_{{\forall l_{j}\in\mathcal{L}}} (w_i+w_j)\mathcal{E}_{(i,j)} \text{ where } i>j \nonumber \\
    &=&\sum_{{\forall l_{i}\in\mathcal{L}}} \sum_{{\forall l_{j}\in\mathcal{L}}} (w_i+w_j)(\mathcal{I}_{i}\land \mathcal{I}_{j}) \text{ where } i>j.
\end{eqnarray}
Here, $\land$ is a Boolean AND operator.
Due to quantum Fourier expansion, the AND Boolean function $B_2(x_1,x_2)$ can be mapped to the following Boolean Hamiltonian $H_{B_2}$:
%
\begin{eqnarray}
    B_2(x_1,x_2)&=&x_1\land x_2 \text{ where }\nonumber \\
    & & x_1\in \{0,1\}\text{ and } x_2\in \{0,1\}, \label{eq:boolean_f2} \\
    H_{B_2}&=&\frac{1}{4}(I-Z_1-Z_2+Z_1 Z_2), \label{eq:boolean_H2}
\end{eqnarray}
where $Z_{1}$ and $Z_{2}$ are the Pauli-$Z$ operators applying on $x_1$ and $x_2$, respectively.

According to~\eqref{eq:boolean_f2}--\eqref{eq:boolean_H2}, the constraint function~\eqref{eq:C} can be represented as following Hamiltonian:
%
\begin{equation}
    H_{C'}=  \sum_{{\forall l_{i}\in\mathcal{L}}} \sum_{{\forall l_{j}\in\mathcal{L}}}  \frac{1}{4}(w_i+w_j)(I-Z_i-Z_j+Z_i Z_j) \text{ where } i>j.
\end{equation}
Here, $Z_{i}$ and $Z_{j}$ are the Pauli-$Z$ operators applying on $\mathcal{I}_i$ and $\mathcal{I}_j$, respectively.
The constraint of the model is to minimize $H_{C'}$, and then the constraint Hamiltonian $H_C$ is as follows:
%
\begin{equation}\label{eq:H_C}
    H_{C}=\sum_{{\forall l_{i}\in\mathcal{L}}} \sum_{{\forall l_{j}\in\mathcal{L}}}-\frac{1}{4}(w_i+w_j)(Z_i+Z_j-Z_i Z_j) \text{ where } i>j.
\end{equation}

Based on the definitions of $H_O$ and $H_C$, the problem Hamiltonian $H_P$ can be defined as follows:
%
\begin{equation}\label{eq:H_P}
    H_P=H_O+\rho H_C, 
\end{equation}
where $\rho \in\mathbb{R}^{+}$ is the penalty rate, which indicates the rate of which $H_C$ affects $H_P$ compared to $H_O$.

\subsection{Design the mixing Hamiltonian}
The mixing Hamiltonian, denoted by $H_{M}$, generates a variety of cases that can appear in the problem.
MWIS can be formulated by a binary bit string that represents a set of nodes (e.g., $\ket{1010101}$); thus various cases can be created by flip the state of each node represented by $\ket{0}$ or $\ket{1}$.
The bit-flip can be handled by the Pauli-$X$ operator, thus $H_{M}$ is as follows:
%
\begin{equation}\label{eq:H_M}
    H_M=\sum_{{\forall l_{k}\in\mathcal{L}}} X_k,
\end{equation}
where $X_{k}$ is the Pauli-$X$ operator applying on $\mathcal{I}_{k}$.
In other words, $H_M$ is a transverse-field Hamiltonian.

\subsection{Apply to QAOA sequence} 
The application of the designed Hamiltonian to QAOA sequence starts to conduct when the design of Hamiltonian, i.e., $H_P$ and $H_M$, is completed. 
First, the parameterized state $\ket{\gamma, \beta}$ can be generated by applying $H_P$ and $H_M$ defined in~\eqref{eq:H_O},~\eqref{eq:H_C},~\eqref{eq:H_P}, and~\eqref{eq:H_M}, to~\eqref{eq:state}.
Here, the initial state $\ket{s}$ is set to the equivalent superposition state using the Hadamard gates.
The expectation value of $H_P$ can be measured on the generated parameterized state $\ket{\gamma, \beta}$.
The $2p$ parameters $\gamma$ and $\beta$ are iteratively updated in a classical optimization loop.
When the QAOA sequence terminates, the optimal parameters $\gamma_{opt}$ and $\beta_{opt}$ are obtained.
Thus the scheduling solution can be obtained by the measurement of the expectation value of $H_P$ on the optimal state $\ket{\gamma_{opt}, \beta_{opt}}$ as follows:
%
\begin{equation}
    \expval{F}=\expval{H_P}{\gamma_{opt}, \beta_{opt}},
\end{equation}
where $\expval{F}$ is the expectation value of the objective function~\eqref{obj_f} over the returned solution samples.

\section{Performance Evaluation}\label{sec:5}
The proposed QAOS algorithm is implemented using Cirq and TensorFlow-Quantum developed for NISQ algorithm and quantum machine learning computation~\cite{broughton}.

\subsection{Software Implementation}
The application of the quantum gates, the basic units of the quantum circuit, is expressed by unitary operators.
Based on the definitions of Hamiltonians in Section~\ref{sec:4}, the objective operator $U_O(\gamma_{\zeta})$, constraint operator $U_C(\gamma_{\zeta})$, problem operator $U_P(\gamma_{\zeta})$, and mixing operator $U_M(\beta_{\zeta})$ which are unitary operators can be defined as follows:
%
\begin{eqnarray}
    U_O(\gamma_{\zeta})&=&e^{-i\gamma_{\zeta} H_O}, \label{eq:U_O}\\
    U_C(\gamma_{\zeta})&=&e^{-i\gamma_{\zeta} \rho H_C}, \label{eq:U_C}\\
    U_P(\gamma_{\zeta})&=&U_O(\gamma_{\zeta}) U_C(\gamma_{\zeta}) = e^{-i\gamma_{\zeta} (H_O+\rho H_C)}, \label{eq:U_P}\\
    U_M(\beta_{\zeta})&=&e^{-i\beta_{\zeta} H_M}, \label{eq:U_M}
\end{eqnarray}
where $\gamma_{\zeta}$ and $\beta_{\zeta}$ are in $\gamma\equiv\gamma_{1}\cdots\gamma_{p}$ 
and $\beta\equiv\beta_{1}\cdots\beta_{p}$, respectively: $\zeta \in\mathbb{Z}^{+}$ and $1\leq \zeta \leq p$.
Note that implementing $U_P(\gamma_{\zeta})$ and $U_M(\beta_{\zeta})$ is the core of QAOS implementation. 

%
\begin{figure}[t]
    \begin{lstlisting}
# - QUBO Model for MWIS
def get_MWIS_QUBO(graph: nx.Graph, penalty_rate=1):
    weight_set = np.array(
        [graph.nodes(data=True)[i]['weight'] for i in range(graph.number_of_nodes())])
    problem_QUBO={}
    for i in range(graph.number_of_nodes()):
        problem_QUBO[(i,)]=weight_set[i]
        for _,j in graph.edges(i):
            problem_QUBO[(i,)]-=(weight_set[i]+weight_set[i])*penalty_rate/2
            if i < j:
                problem_QUBO[(i, j)]=(weight_set[i]+weight_set[i])*penalty_rate/2
    return problem_QUBO
...
# - Problem Operator
def problem_operator(p_QUBO:dict, qubits, p, gamma):
    key_iter=sorted(p_QUBO.keys(), key=lambda x: (len(x), x))
    for nodes in key_iter:
        for i in range(len(nodes)-1):
            yield cirq.CNOT(qubits[nodes[i]], qubits[nodes[i+1]])
        yield cirq.rz(gamma[p]*p_QUBO[nodes])(qubits[nodes[-1]])
        for i in range(len(nodes)-1):
            yield cirq.CNOT(qubits[nodes[i]], qubits[nodes[i+1]])
...
# - Mixing Operator   
def mixing_operator(mwis_graph, qubits, p, beta):
    for node in mwis_graph.nodes:
        yield cirq.rx(2*beta[p])(qubits[node])
...
# - Optimal Parameter Search using Keras Model
model = tf.keras.Sequential()
model.add(tf.keras.layers.Input(shape=(), dtype=tf.dtypes.string))
model.add(tfq.layers.PQC(model_circuit, model_readout)) # Parameterized Quantum Circuit
model.add(tf.keras.layers.Lambda(correction))
model.compile(loss=tf.keras.losses.mean_absolute_error,
              optimizer=tf.keras.optimizers.Adam(learning_rate=learning_rate))
history = model.fit(input_,optimum,epochs=epochs,verbose=0)
\end{lstlisting}
    \caption{Parts of Python codes using Cirq and TensorFlow-Quantum for solving the MWIS-based scheduling problem.}
    \label{fig:code}
\end{figure}
In Fig.~\ref{fig:code}, \texttt{cirq.rz()} and \texttt{cirq.CNOT()} are used for the implementation of $U_P(\gamma_{\zeta})$.
Note that, \texttt{cirq.rz()} and \texttt{cirq.CNOT()} represent the rotation-$Z$ gate and the controlled-NOT gate, respectively.
In addition, $U_M(\beta_{\zeta})$ is implemented using \texttt{cirq.rx()} which represents the rotation-$X$ gate.

The part that finds the optimal parameters using Keras (one of the well-known open-source deep learning computation libraries) is shown in Fig.~\ref{fig:code} from line $29$ to line $36$.
Here, the parametrized quantum circuit (PQC) layer provides auto-management of variables in the parameterized circuit.
In this model, Adam is used as a gradient-based optimizer~\cite{zhang, kingma}. 

\subsection{Experimental Results} 
The performance of proposed QAOS algorithm is compared with the random search and greedy search~\cite{feo}. 
In addition, the QAOS algorithm executes with different $p$ value settings where the $p$ value means the number of alternations of $U_P(\gamma_{\zeta})$ and $U_M(\beta_{\zeta})$ in~\eqref{eq:U_P} and~\eqref{eq:U_M}, i.e., $\zeta \in\mathbb{Z}^{+}$ and $1\leq \zeta \leq p$.

\begin{figure}[ht]
    \centering
        \includegraphics[width =0.77\columnwidth]{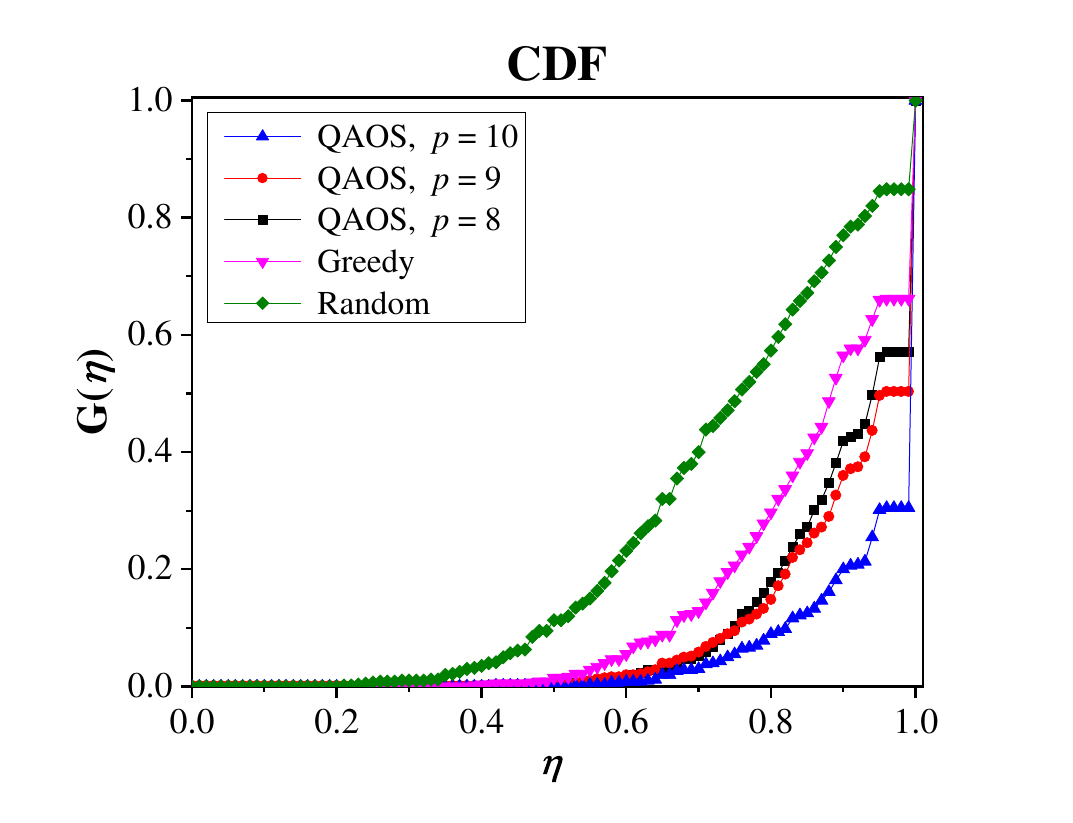}
        \vspace{-2mm}
    \caption{Performance Evaluation Results. $G(\eta)$ is the cumulative distribution function (CDF) of $\eta$.}
    \label{fig:sim}
\end{figure}

For the performance evaluation, random conflict graphs with $10$ nodes are generated; and then random search, greedy search, and QAOS algorithms are performed for the given random conflict graphs.
The measurement of each QAOS is performed $1,000$ times in each simulation (i.e., in each randomly generated conflict graph).
The performance of each algorithm is quantitatively measured with $\eta$ as follows:
\begin{equation}
    \eta\triangleq \frac{a}{b},
\end{equation}
where $a$ and $b$ are the summations of weights of the scheduled nodes by the used algorithms and the summations of weights of the scheduled nodes by brute-force search (i.e., exhaustive search), respectively, for the given randomly generated graphs.
As shown in Fig.~\ref{fig:sim}, the cumulative distribution functions (CDF) of $\eta$ for each algorithm are computed.

\begin{table}[ht]
\vspace{4mm}
\caption{Percentage of Optimal Solution Computation.}
\label{tab:tab1}
\normalsize
\vspace{-3mm}
\begin{center}
	\centering
	\begin{tabular}{c|c|c|c|c}
    \toprule[1.0pt]
    \centering
    QAOS, $p=10$ & QAOS, $p=9$ & QAOS, $p=8$ & Greedy & Random \\
    \midrule[1.0pt]
    $69.50$\% & $49.67$\% & $42.83$\% & $33.83$\% & $15.17$\% \\
    \bottomrule[1.0pt]
	\end{tabular}
\end{center}
\end{table}
As presented in Fig.~\ref{fig:sim}, QAOS algorithms with $p\geq 8$ present better performance than random search and greedy search, in any kinds of randomly generated conflict graphs.
In these repeated simulations, the performances of QAOS algorithms are improved as $p$ value increases.
In particular, the performance of QAOS algorithm with $p=10$ is much better than the QAOS algorithms with $p=8 $ and $p=9$.
As shown in Table~\ref{tab:tab1}, the QAOS algorithm with $p=10$ returns optimal solutions (i.e., equivalent to the solutions obtained by brute-force search) with a ratio of $69.50\%$.
Through these, it has verified that the proposed QAOS algorithm presents beautiful results in terms of the accuracy of the solutions.

\section{Concluding Remarks and Future Work}\label{sec:6}
The wireless scheduling can be modeled with the MWIS problem which is one of the well-known NP-hard problems. In order to solve the MWIS problem, a QAOA-based scheduling algorithm, so-called quantum approximate optimization for scheduling (QAOS), is proposed. 
The proposed QAOS is implemented using Cirq and TensorFlow-Quantum.
QAOS outperformed greedy search and random search in the performance evaluation on the random conflict graphs.
Therefore, the quantum approach to the wireless scheduling problem using QAOS is effective in terms of accuracy.

Future research focuses on improving the performance of QAOS.
In one method, introducing an error correction code to QAOS is considered.
This method is expected to improve the sampling quality.
Another method is to develop a new optimizer that can more accurately find the optimal parameters of QAOS.
A novel optimizer is needed that is more suitable for quantum models than the mainly used optimizers such as Adam, Nelder–Mead (NM), and Broyden-Fletcher-Goldfarb-Shanno (BFGS).
From the perspective of quantum machine learning, developing the novel optimizer for the parameterized quantum circuit like the QAOS circuit will be a meaningful challenge. 

\bibliographystyle{IEEEtran}

\end{document}